\documentstyle[aps,pra,epsf,multicol]{revtex}
\begin{document}
\draft
\title{Purification via entanglement swapping and conserved entanglement}
\author{S. Bose $^1$, V. Vedral $^2$ and P. L. Knight $^1$}
\address{$^1$Optics Section, The Blackett Laboratory, Imperial College, London SW7 2BZ, England}
\address{$^2$ Centre for Quantum Computing, Clarendon Laboratory,
	University of Oxford,
	Parks Road,
	Oxford OX1 3PU, England}

\maketitle
\begin{abstract}
We investigate the purification of entangled states
by local actions using a variant of entanglement swapping. 
We show that there exists a measure of entanglement
which is conserved in this type of purification procedure. 
\end{abstract}

\begin{multicols}{2}

The resource of entanglement \cite{revs} has many
useful applications in quantum
information processing, such as secret key distribution 
\cite{Ek}, teleportation \cite{ben} and dense coding \cite{wies}. Polarization entangled photons have been used to
demonstrate both dense coding \cite{zei1} and teleportation
\cite{zei2} in the laboratory. 
Teleportation has also been realized using path-entangled photons \cite{hrd} and entangled
electromagnetic field modes \cite{braun}. Accompanying the practical
applications of entanglement are some useful schemes for
entanglement manipulation which
may help in the distribution of entanglement
between distant parties. One such scheme is entanglement 
swapping \cite{zei3,us}, which enables one to entangle two
quantum systems that have never interacted directly with
each other; we have discussed how this may be used in constructing a quantum
telephone exchange \cite{us}. Recently,
entanglement swapping has been demonstrated  
experimentally \cite{zei4}. There exists yet another useful manipulation of
entanglement in
which local actions and classical communication are used 
by two distant parties to distill a certain number of shared Bell states
from a larger number of shared but less entangled states. When the
initial shared but less entangled states are pure, this manipulation 
is termed as entanglement concentration \cite{pf1,lo}, while for
the more general case of the initial shared states being mixed the
process is termed as entanglement purification or distillation \cite{mix,pf2}.
The importance
of such a scheme in the distribution of entanglement is obvious as 
Bell pairs are
essential for the implementation of quantum communication schemes 
with perfect fidelity.  
Curiously, 
such an important 
procedure remains 
to be realized in an experiment. In this paper, we will show that a 
simple variant of
the entanglement swapping scheme can be viewed as a type of entanglement
concentration procedure and has a physical realization with
polarization entangled photons. Moreover, we 
show that there exists a certain measure of entanglement
which remains conserved on average in this type of
entanglement concentration. Note that, in this paper we will often use
the terms entanglement concentration and entanglement purification in an
interchangeble manner, though what we demonstrate is, in the strict
sense, only the concentration of pure shared entanglement.

\begin{figure}
\begin{center} 
\leavevmode 
\epsfxsize=8cm 
\epsfbox{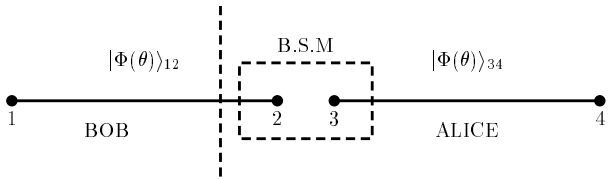}
\vspace{1cm}
\caption{\narrowtext The figure illustrates the procedure of
purification via entanglement swapping. The solid lines connect
initially entangled photon pairs. The dashed line divides the
set of photons to those held by Alice and those held by Bob. B.S.M
denotes the local Bell state measurement done by Alice.}
\label{swpr} 
\end{center}
\end{figure}

         Let pairs of 
photons $(1,2)$ and $(3,4)$ be in the following 
polarization entangled states
\begin{mathletters} 
\begin{eqnarray}
\label{ampl1}
 |\Phi (\theta)\rangle_{12} = \cos{\theta}|H_1,H_2\rangle + \sin{\theta}|V_1,V_2\rangle,  \\
|\Phi (\theta)\rangle_{34} = \cos{\theta}|H_3,H_4\rangle + \sin{\theta}|V_3,V_4\rangle,
\label{ampl2}
\end{eqnarray}
\end{mathletters}
where the {\em phase angle}
$\theta$ satisfies $0 < \theta < \pi/2$. There are a
number of ways to prepare photons in such polarization
entangled states. For example, one may first use type-II down conversion
followed by suitable birefringent crystals to prepare two photons
in the Bell state
state $|\Phi\rangle^{+}=|HH\rangle + |VV\rangle$ \cite{zei5} (the other
Bell states being $|\Phi\rangle^{-}=|HH\rangle - |VV\rangle$ and
$|\Psi\rangle^{\pm}=|HV\rangle \pm |VH\rangle$). This may be followed
by placing a dichroic element (such as the local filters 
described in Ref.\cite{gis}) which selectively absorbs any one of the  
polarization components (say $H$) along the path of one of the photons.
In cases
when this photon exits the element unabsorbed, the pair of
photons are in
the state $e^{-\gamma L}|HH\rangle \pm |VV\rangle$ (not normalized), 
where $L$ is the
length of the crystal and $\gamma$ is the absorption per unit
length. Thereby states of the type given by Eqs.(\ref{ampl1})
and (\ref{ampl2}) with $\sin{\theta}=\sqrt{1/(1+e^{-2\gamma L})}$ is generated.
This procedure may seem inefficient because of the possibility of the
photon being absorbed by the dichroic element. However, due to the absence of two qubit logic gates for polarization entangled photons, there is no way to proceed unitarily from $|HH\rangle\pm|VV\rangle$ to the states given by Eqs.(\ref{ampl1})
and (\ref{ampl2}) and dissipative processes are necessary. Alternatively, one can use the recently suggested tunable
ultrabright source of polarization entangled photons \cite{kw} to
directly produce the states given in Eqs.(\ref{ampl1})-(\ref{ampl2}).  
Photons $2$ and $3$ are brought together, while
photons $1$ and $4$ are allowed to travel to separate distant locations
as shown in Fig.\ref{swpr}. If one now performs a
Bell state measurement on photons $2$ and $3$, then
immediately the states of the photons $1$ and $4$ become
entangled. This effect, called entanglement swapping, has been
tested for maximally entangled photons (when $\theta=\pi/4$) \cite{zei4}.
We shall refer to this tested case (i.e when maximally entangled photons
are used) as standard entanglement swapping.  
For the more general case of an arbitrary $\theta$, the 
combined state of the photons
$1$ and $4$ is projected to either of the following four states, 
depending on the outcome of the Bell state measurement
on photons $2$ and $3$:
     
\begin{mathletters} 
\begin{eqnarray}
\label{states1} 
|\Phi^{'}(\theta)\rangle_{14}^{+}&=&\frac{1}{N}(\cos^2{\theta}|H_1,H_4\rangle + \sin^2{\theta}|V_1,V_4\rangle),  \\
|\Phi^{'}(\theta)\rangle_{14}^{-}&=&\frac{1}{N}(\cos^2{\theta}|H_1,H_4\rangle - \sin^2{\theta}|V_1,V_4\rangle), \\
|\Psi\rangle_{14}^{+}&=&\frac{1}{\sqrt{2}}(|H_1,V_4\rangle+|V_1,H_4\rangle), \\
|\Psi\rangle_{14}^{-}&=&\frac{1}{\sqrt{2}}(|H_1,V_4\rangle-|V_1,H_4\rangle),
\label{states2}
\end{eqnarray}
\end{mathletters}
where, $N=\sqrt{\cos^4{\theta}+\sin^4{\theta}}$. The probabilities to obtain the four states are 

\begin{mathletters} 
\begin{eqnarray}
\label{prob1}
P(|\Phi^{'}(\theta)\rangle_{14}^{+})
&=& P(|\Phi^{'}(\theta)\rangle_{14}^{-}) = \frac{\cos^4{\theta}+\sin^4{\theta}}{2}, \\
P(|\Psi\rangle_{14}^{+}) &=& P(|\Psi\rangle_{14}^{-}) = \cos^2{\theta} \sin^2{\theta},
\label{prob2}
\end{eqnarray}
\end{mathletters}
where the symbol $P(|\psi\rangle)$ is used to denote the probability
of the state $|\psi\rangle$.
We now proceed to explain the sense in which the above variant of
standard entanglement swapping is a purification procedure. Say photons
$2$,$3$ and $4$ are held by one party (Alice) and photon
$1$ is possessed by the other party (Bob) as shown
in Fig.\ref{swpr}. Now, Alice can change the magnitude of the entanglement
she shares with Bob by doing a Bell state measurement on
photons $2$ and $3$ and thereby projecting photons $1$
and $4$ to one of the states given by Eq.(\ref{states1})-(\ref{states2}).
In the cases when she projects photons $1$ and $4$ to
either $|\Psi\rangle_{14}^{+}$ or $|\Psi\rangle_{14}^{-}$ (which are
Bell states), she actually
increases the magnitude of entanglement she shares with
Bob. In the other cases she reduces the magnitude of
entanglement she shares with Bob even further. If she initially shared
a large enough ensemble of photons in the state $|\Phi (\theta)\rangle_{12}$
with Bob and applied the procedure
described above on each shared pair separately, then she would be
able to change the states of a certain
fraction of the shared pairs to Bell states at the expense of degrading the entanglement of the other shared pairs even
further. This qualifies as a type of purification procedure because local 
actions (by Alice) are used to concentrate the
entanglement of a fraction of the
shared states while the entanglement
of the remaining fraction is being diluted, just as in other purification
procedures \cite{pf1,pf2}.

\vspace{2cm}

         We should pause here briefly to mention a fact relevant
from the experimental viewpoint. It is known that a complete
Bell state measurement is not possible with only linear elements \cite{lut}.
However, a complete Bell state measurement is {\em not} really {\em necessary} to
implement the above purification protocol. One only needs to
be able to discriminate the states $|\Psi\rangle_{23}^{+}$ and
$|\Psi\rangle_{23}^{-}$ from each other and from the rest of the
Bell states, which can be done in existing experiments \cite{zei1,zei2,zei3,zei4}. The reason for this is that the photons $1$ and $4$
are projected to Bell states ($|\Psi\rangle_{14}^{+}$ or $|\Psi\rangle_{14}^{-}$) 
only when the outcome of the Bell state
measurement on photons $2$ and $3$ is either $|\Psi\rangle_{23}^{+}$ or
$|\Psi\rangle_{23}^{-}$. To verify whether the purification has indeed
taken place, one simply has to pass the resultant Bell states of photons
$1$ and $4$ through a Bell state analyser \cite{zei1,zei2,zei3,zei4}.

         An interesting feature of the original collective entanglement concentration
procedure (called the "Schmidt projection method") described in Ref.\cite{pf1} was that the average
of the entropy of entanglement (the von Neumann entropy of the
partial density matrix seen by either party) of all the shared pairs 
remained constant under purification (an asymptotic
result). In the scheme described here,
the average von Neumann entropy of entanglement of the
shared pairs is, in fact, decreased. However, 
we shall show that there is a different measure of
entanglement whose average remains conserved in this
procedure of purification via entanglement swapping. This measure of entanglement is defined as the maximum probability with which
two parties sharing a pure entangled state can convert it 
to a Bell state by classically communicating and
performing local actions on their respective
sides. As this is already the maximum
probability, it can only remain constant or decrease under
any set of local general measurements and classical communications
(This is the basic criterion to be satisfied by any measure
of entanglement \cite{mix,vlat}).  
Thereby, it qualifies as a measure of entanglement which
can be termed as {\em entanglement of single pair purification} (This
measure, is however, not an additive measure). 
We shall denote this measure by
$E_S$ and refer to it henceforth simply as entanglement.

              We now proceed
to show that $E_S$ is conserved
in the purification process described
above. From the results of Lo and Popescu \cite{lo}, it 
follows that the maximum probability with which a Bell state can be
obtained by purifying a single entangled pair (in a pure state) is twice the
modulus square of the Schmidt coefficient of smaller magnitude. In the
case of the states given by Eqs.(\ref{ampl1}) and (\ref{ampl2}), this is
simply $2\cos^2{\theta}$ if $\cos{\theta}$ is the smaller of the
two Schmidt coefficients. Therefore, before the entanglement
swapping, the average value of the entanglement shared
between Alice (A) and Bob (B) is
\begin{equation}
\langle E_S \rangle_{AB} = 2\cos^2{\theta}.
\end{equation}
After the entanglement swapping, when the shared states are those
given by Eqs.(\ref{states1})-(\ref{states2}) with probabilities
given by Eqs.(\ref{prob1})-(\ref{prob2}),

\begin{eqnarray}
\langle E_S \rangle_{AB} &=& P(|\Phi^{'}(\theta)\rangle_{14}^{+}) E_S(|\Phi^{'}(\theta)\rangle_{14}^{+})\nonumber \\ &+& P(|\Phi^{'}(\theta)\rangle_{14}^{-}) E_S(|\Phi^{'}(\theta)\rangle_{14}^{-}) \nonumber \\
&+& P(|\Psi\rangle^{+}) E_S(|\Psi\rangle^{+}) + P(|\Psi\rangle^{-}) E_S(|\Psi\rangle^{-}) \nonumber \\
&=& 2\cos^2{\theta},
\end{eqnarray}
where $E_S(|\psi\rangle)$ denotes the entanglement of the state $|\psi\rangle$.
Therefore, the ensemble average of the entanglement of single pair purification is a conserved
quantity in the process of purification by entanglement swapping. This result
also indicates that the purification via entanglement swapping is an
 optimal protocol for single pair entanglement
 purification, as the average entanglement of the purified pairs is equal to
 the original entanglement. Here, by optimality we mean the best combination
 of entangled states that can be finally obtained by the purification.
 A positive implication of this optimality is that the purification
 process can be
 separately repeated on the final subensembles, which turn out to be less 
 entangled, without destroying any entanglement on average (though
 for that, one needs to be able to do complete Bell state measurements, for
 which schemes have been suggested \cite{kw1}).
 If one continues this process indefinitely, in the limit of an infinite
 sequence, the final ensemble generated will comprise of a certain fraction
 of Bell pairs and a certain fraction of completely disentangled pairs. 
 This fraction of Bell pairs should be equal to $2\cos^2{\theta}$ as the
average of the entanglement has been conserved in each step. Thus
 in the limit of an infinite sequence, the process of purification
 via entanglement swapping allows us to convert all the entanglement that
 can possibly be extracted by single pair purifications from the ensemble
 into Bell pairs.
\vspace{2cm}
\begin{figure}
\begin{center} 
\leavevmode 
\epsfxsize=8cm 
\epsfbox{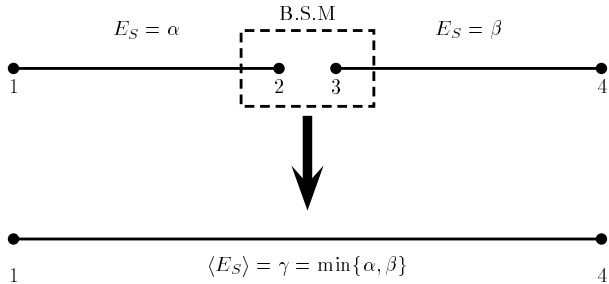}
\caption{\narrowtext The figure shows the generalization of entanglement swapping
to pure states with arbitrary entanglement.}
\label{gen} 
\end{center}
\end{figure}

   Here, it is worthwhile to mention that a procedure for single
pair purification called the 
"Procrustean method" had also been suggested in the
original entanglement concentration paper \cite{pf1}. The  Procrustean
method also conserves $E_S$ and its fractional yield of Bell states is optimal
(i.e equal to $E_S$). From the point of view of efficiency, our method
lies somewhere between the Schmidt projection method (when it is
implemented with two entangled pairs, but one pair totally belonging to Alice)
and the procrustean method. Our method has a fractional yield of Bell states equal to the
Schmidt projection method (with two entangled pairs), but the non-Bell state
outcomes are also entangled. It is because of the extra entanglement of the
non-Bell state
outcomes that $E_S$ is conserved on average in our method of purification.
Hence the improvement over the Schmidt projection method, brought in by
 entanglement swapping is {\em to make the non-Bell state
outcomes entangled as well}.

   Next, we proceed to consider the case when photon pairs $(1,2)$ and
$(3,4)$ are not in the same type of entangled state. Suppose, photons 
$1$ and $2$ start in the entangled state $|\Phi (\theta_1)\rangle_{12}$
and photons $3$ and $4$ start in the entangled state $|\Phi (\theta_2)\rangle_{34}$ (defined as in Eqs.(\ref{ampl1}) and 
(\ref{ampl2})). Let the entanglement of the first photon pair be $E_S=\alpha$
and the entanglement of the second photon pair be $E_S=\beta$.
Then a simple calculation shows that projecting photons $2$ and $3$ onto a Bell state basis projects the photons $1$ and $4$ to states
$(\cos{\theta_1}\cos{\theta_2}|H_1,H_4\rangle \pm \sin{\theta_1}\sin{\theta_2}|V_1,V_4\rangle)$ and $(\cos{\theta_1}\sin{\theta_2}|H_1,V_4\rangle \pm \sin{\theta_1}\cos{\theta_2}|V_1,H_4\rangle)$ (not normalized) with
specific probabilities.
 The average of the entanglement between the photons $1$ and $4$ after the projection
turn out to be $\langle E_S \rangle = \mbox{min}\{ \alpha, \beta \}$.
  The manipulation of entanglement described above can be visualized
as a step towards the complete generalization of entanglement swapping.
The original scheme involved the use of two Bell states and
Bell state measurements \cite{zei3,zei4}. 
It has been generalized to cases when many 
particle Greenberger-Horne-Zeilinger (GHZ) states are
used and many particle GHZ measurements are conducted \cite{us}. The
procedure presented here is a generalization of the original scheme to
the case when two particle pure states which are less entangled than Bell
states are used, but measurements conducted are still Bell state 
measurements. 
This result is illustrated in Fig.\ref{gen}. One can get an intuitive
feel of this result from the golden rule that entanglement, on 
average, cannot be increased under local actions and classical
communications \cite{vlat}. In the situation depicted
in Fig.\ref{gen}, we have the freedom to choose which photons
belong to Alice and which to Bob. For $\alpha < \beta$,  we
choose to allot the photon $1$ to Bob and the rest to Alice, while
for $\alpha > \beta$ we allot the photon $4$ to Bob and the rest to Alice.
Then a Bell state measurement on Alice's side cannot increase the
entanglement she shares with Bob on average. As this initial
entanglement is the smaller of the numbers $\alpha$ and $\beta$, the
final average entanglement has to be smaller than or equal to
$\mbox{min}\{ \alpha, \beta \}$. However, the fact that it is actually equal,
is a peculiar feature of the entanglement swapping process.

        One may envisage a
situation in which Alice tries to purify the state 
$|\Phi (\theta_1)\rangle_{12}$ shared with Bob with the help of the
state $|\Phi (\theta_2)\rangle_{34}$ (which we can call the
purifier) in her possession. As long as
$\beta < \alpha $, she degrades the entanglement shared with Bob on
average, while when $\beta \geq \alpha$, she conserves
the entanglement shared with Bob on
average. Thus in order not to loose any entanglement the purifier
state should have at least as much entanglement as the state to
be purified. Moreover, the degree to which the entanglement
is concentrated (i.e inhomogeneously redistributed among the
four measurement outcomes) gets better as $\beta$ approaches $\alpha$.
 Bell pairs are produced {\em only} when $\beta$ {\em exactly
equals} $\alpha$. Thus, as Alice increases the
entanglement of her purifier, the degree of
entanglement concentration increases yielding Bell pairs
when $\beta$ reaches $\alpha$ (a criterion which
can be called {\em entanglement matching}). On increasing 
 $\beta$ further, the degree of concentration starts going
down. When $\beta$ reaches unity (maximal entanglement), there
is no concentration of entanglement (all the four outcomes
have an entanglement equal to that of the original state: a situation
equivalent to the perfect fidelity teleportation of an entangled
state).          
             
                In this paper we have shown that entanglement
                swapping can be used to purify single pairs of
                polarization entangled photons. Such a scheme may
be achieved by an extension of an existing experiment \cite{zei4}. In contrast, 
physical implementation of the entanglement purification schemes
involving collective measurements (like the Schmidt
projection method of Ref.\cite{pf1}) will be difficult, as they would involve
measurements on many photons at once. The absence of two qubit logic gates for polarization entangled photons also make purification procedures
described in Ref.\cite{pf2} difficult to realize. From this point of
view, the method described here should be a positive first step
in the direction of implementation of purification procedures. In the
paper we have also
                introduced a measure of entanglement for single
                pure pairs and demonstrated that this quantity
                is conserved in the above process. This makes
purification via entanglement swapping interesting from a
purely theoretical angle. The non-additive
nature of the introduced measure implies that additivity is not an essential
requirement for an entanglement measure to have a physical interpretation.
As a natural generalization of our measure one may use the maximum 
possible fractional yield of
Bell states from various {\em finite} collections of shared
pairs as physically relevant measures of entanglement. The physical
relevance stems from the fact that in a real implementation of entanglement concentration one would
always have access to only a finite number of systems. 
All the results presented in this paper hold only for pure states. An 
extension to mixed states will not be trivial, as single pairs in
such states cannot, in general, be purified \cite{knt}.
However, it would still be interesting to investigate whether one can generalize
the measure of entanglement presented here to some quantity 
which is conserved in entanglement
swapping with mixed states.            
                
            This work was supported in part by the Inlaks Foundation, the UK Engineering and Physical Sciences Research Council, the European Union, 
Elsag-Bailey and 
Hewlett-Packard.

\end{multicols} 

\end{document}